\documentclass[11pt,fleqn]{article}   

\usepackage{latexsym,bm}
\usepackage{amsmath}
\usepackage{amsfonts,amssymb}
\usepackage{cite}

\catcode`\@=11
\@addtoreset{equation}{section}
\def\theequation{\arabic{section}.\arabic{equation}}
\def\appendix{\renewcommand{\thesection}{\Alph{section}}\setcounter{section}{0}
              \renewcommand{\theequation}
            {\mbox{\Alph{section}.\arabic{equation}}}\setcounter{equation}{0}}

\def\maketitle{\thispagestyle{empty}\setcounter{page}0\newpage
                \renewcommand{\thefootnote}{\arabic{footnote}}
                  \setcounter{footnote}0}
\renewcommand{\thanks}[1]{\renewcommand{\thefootnote}{\fnsymbol{footnote}}
               \footnote{#1}\renewcommand{\thefootnote}{\arabic{footnote}}}

\renewcommand{\title}[1]{\begin{center}\Large\bf #1\end{center}\rm\par\bigskip}
\renewcommand{\author}[1]{\begin{center}\Large #1\end{center}}

\newcommand{\pacs}[1]{\smallskip\noindent{\sl PACS numbers:
                       \hspace{0.3cm}#1}\par\bigskip\rm}
\def\babs{\hrule\par\begin{description}\item{Abstract: }\it} 
\def\eabs{\par\end{description}\hrule\par\medskip\rm}
\renewcommand{\date}[1]{\par\bigskip\par\sl\hfill #1\par\medskip\par\rm}
 
\newcommand{\s}[1]{\section{#1}}

\newcommand{\ca}[1]{{\cal #1}}         
\def\nn{\nonumber}            
\def\beq{\begin{eqnarray}}    
\def\eeq{\end{eqnarray}}      
\def\ben{\begin{itemize}}    
\def\een{\end{itemize}}      
\def\at{\left(}               
\def\aq{\left[}               
\def\ag{\left\{}              
\def\ct{\right)}              
\def\cq{\right]}              
\def\cg{\right\}}             
\def\R{{\hbox{{\rm I}\kern-.2em\hbox{\rm R}}}}   
\def\H{{\hbox{{\rm I}\kern-.2em\hbox{\rm H}}}}   
\def\N{{\hbox{{\rm I}\kern-.2em\hbox{\rm N}}}}   
\def\C{{\ \hbox{{\rm I}\kern-.6em\hbox{\bf C}}}} 
\def\Z{{\hbox{{\rm Z}\kern-.4em\hbox{\rm Z}}}}   
\def\ii{\infty}                                  
\def\pa{\partial}
\def\al{\alpha}
\def\be{\beta}
\def\ga{\gamma}
\def\de{\delta}
\def\ep{\varepsilon}

\def\ka{\kappa}
\def\la{\lambda}

\def\si{\sigma}
\def\om{\omega}

\def\La{\Lambda}
\def\Si{\Sigma}

\def\o{\perp}
\def\p{\parallel}

\begin{document}
\maketitle
\hfill{\sl preprint - UTM 656/UTF 453} 
\par 
\bigskip 
\par 
\rm

 
\par 
\bigskip 
\LARGE 
\noindent 
{\bf  Central Charges and Boundary Fields for Two Dimensional
  Dilatonic Black Holes}  
\bigskip 
\par 
\rm 
\normalsize 
 
  
\large 
\noindent {\bf Nicola Pinamonti$^{1}$ and Luciano Vanzo$^{2,3}$}

\noindent 
$^1$ Dipartimento di Matematica, E-mail: pinamont@science.unitn.it\\ 
$^2$ Dipartimento di Fisica, E-mail: vanzo@science.unitn.it\\ 
$^3$ I.N.F.N. Gruppo Collegato di Trento\\ 
\rm\large \large 
 
\noindent 
 Universit\`a di Trento,\\ 
 Facolt\`a di Scienze,\\
Via Sommarive 14,\\ 
I-38050 Povo (TN),\\ Italia. 
\large 
\smallskip 
 
\rm\normalsize 
 
\date{December 2003}

\babs
In this paper we first show that within the Hamiltonian description
  of general relativity, the central charge of a near horizon
asymptotic symmetry group is zero, and therefore that the entropy of
 the system cannot be estimated using Cardy's formula. This is done by
  mapping a static black hole to a two dimensional space. We
  explain how such a charge can only appear to a static observer who
  chooses to stay permanently outside the black hole. Then an alternative
argument is given for the presence of a universal central charge.
Finally we suggest an effective quantum theory on the horizon that is
  compatible with the thermodynamics behaviour of the black hole.  
\eabs
\s{Introduction}

Following the discovery that black holes behave like thermodynamic 
systems (see \cite{wald:01} for a review with an exhaustive list of
references), many efforts have been done to unravel the structure of the
underlying microscopic theory. Most of these focused on the event
horizon of the black hole as the place where interesting new physics
is possibly operating, and in fact the ability to obtain
black hole thermodynamics from horizon quantum states has always  been
considered as a strong test to the viability of any model of quantum
gravity. Entropy computations in particular were one of the main 
activity for people working in string theory and quantum gravity.\\
But as has became clear over the years, the
horizons associated to black holes are not the only  ones that
have thermodynamics. The cosmological horizons, 
the distance light can travel between now and the end of time, also 
have the ability to exchange
information with the regions they trap (classically they would not),
and to create a noisy background radiation with a characteristic
thermal spectrum\cite{gibb-hawk:77}. But in view of
the structureless nature of the 
cosmological horizons the counting of the corresponding microstates is
even 
more acute than for black holes (see \cite{jaco:2003} for a recent
review). Nevertheless it has been shown\cite{davies} that the validity
of the generalized second law would be seriously challenged were it
not for the geometric entropy of the cosmological horizon. 

Perhaps one may notice, at this point, that the entropy we are talking
about is of order $\hbar^{-1}$ and is finite, whereas the one-loop
partition function from field theory, the one from which the
entropy could be derived, is of order $\hbar^0$ and ultraviolet
divergent\footnote{We use this term to emphasize the fact that the
  divergence is not associated with the infinite volume of the
  space.}\cite{thooft85}. This indicates that   
back reaction effects must be important and also that the black hole
spectrum must be strictly discrete, with a quantized area.

Bekenstein \cite{bekenstein2}\cite{bekenstein3} first
argued that if the horizon area were quantized, whichever that means,
its eigenvalues would be equally spaced because the area is an
adiabatic invariant. In this picture the area is a sum of cells with
the same number of degrees of freedom per cell. Since then,
a number of authors
\cite{barvi69,mukha86,koga86,mazu87,belli93,schi93,peleg93,magg94,lous95,beke95,kastr96,kastr97,berez97,vaz00,wittenvaz01}
have given arguments for an equally spaced area spectrum. Moreover,
it has been observed\cite{hod:98,kunst:02,dreyer:03} that the spacing
coefficient can be fixed
using the spectrum of the quasi-normal modes of the Schwarzschild's black
hole. Even earlier than this,
J.~York\cite{york83} proposed to quantize the quasi-normal modes of a non
rotating black hole;  using a ``common sense'' quantization of the
modes he obtained results surprisingly close to the (in)famous $A/4$.

Another approach was to describe the invisible hairs by means of
quantum fields around the black hole. In principle there is nothing
exotic about this picture, but the entropy of these fields was found
to be quadratically divergent as a function of the (inverse) distance
from the horizon\cite{thooft85}. 't Hooft observation was that
regularizing with a cutoff of the order of the Planck length, 
gives an entropy satisfying the area law and with the right
magnitude to be identified with the entropy of the black
hole. Notice that without divergences this would not be possible.

The thermal entropy of quantum fields around the black hole can also
be understood as entropy of entanglement \cite{sorkin86}, since a
static observer hanging over the black hole would have to trace over
the hidden degrees of freedom. The fact that a 
cutoff like the Planck scale must be invoked in an otherwise
conventional theory, 
indicates the necessity to take into account quantum gravity
effects\footnote{This has been stressed particularly lucidly, in our
opinion, by `t Hooft\cite{thooft85}, Jacobson\cite{gr-qc/9404039}, and
Susskind-Uglum\cite{prd50-94}.}.

Still another idea was to describe the microscopy of black
holes by counting the number of states of a conformal field theory
living on the boundary of spacetime. Many years ago J.~Brown and
M.~Henneaux \cite{BrownHenn} made this observation by counting
the dimension of the asymptotic symmetry group at conformal infinity
in anti-de Sitter space. It was
later realized by Strominger\cite{andy98} that this counting gave
complete agreement with the Bekenstein-Hawking entropy of the BTZ
black hole\cite{btz92}. Today we understand this success as a
particular case of the AdS$_{d+1}$/CFT$_d$
correspondence\cite{maldacena:1998}, which describes gravity by a dual
conformal field theory on the boundary at infinity of the conformal
compactification of anti-de Sitter space.

Moreover some supersymmetric extremal and near-extremal black holes  
configurations consisting of D-brane matter were also 
accounted for by a conformal field theory living on the
branes\cite{stro:1996,call-mald:1996,kleb:1996}, so in this case the
boundary was near the horizon rather than at infinity.\\
Therefore people tried to obtain a similar performance for the near
horizon symmetry of non extreme, non supersymmetric
configurations\cite{solo98,carlip99,lin99}, but these treatments 
seemed to have some flaw\cite{park99,park00,solov00}. In particular it
has been shown that after fixing the near
horizon symmetry compatibly with certain boundary conditions, the
central charge of the putative CFT vanishes\cite{hotta00,koga00}.

In this paper we shall show, rigorously, that the
central charge of the near horizon symmetry of a two dimensional
black hole vanishes for asymptotic symmetries that preserve the surface
gravity or the horizon area. We stress that such a charge can only
arise for an observer who chooses to stay outside the black hole and
has an event horizon, because
such an observer must use an Hamiltonian associated to a singular
space-time foliation. 

As in the well known example in three
dimensions, the full group of diffeomorphisms in the presence of a
boundary is not a symmetry group of the theory, but we expect that
this symmetry can be recovered in the full microscopic description. 
Then we shall construct a covariant action for
the theory of gravity in two dimensions with boundary. For this
purpose we add some ad hoc fields living on the horizon. 
Then fixing the energy (the mass of the black hole) we argue that these
fields describe the whole black hole (an uncharged one).  
We show that if we count the degrees of freedom of this gauge
invariant theory on the boundary, we recover the value of the
semiclassical entropy. The remainder of the paper is organized as follows:
In Sec.~[2] we show how a two dimensional black hole with dilaton
can be obtained from dimensional reduction of an higher dimensional one
with spherical symmetry, and why such a theory may explain the entropy
of the parent, original black hole.
In Sec.~[3] we present a detailed description of the Hamiltonian
generators of a two dimensional theory with dilaton, from the
viewpoint of an observer who chooses to stay permanently outside a
given black hole. 
In Sec.~[4] we treat the horizon as an inner boundary and show that the  
Hamiltonian generators corresponding to timelike deformations form an
abelian algebra with no central extension. In sec.~[5] we give an
informal argument suggesting that a central charge must be present
anyway. In sec.~[6] we build an action covariant with respect to every
diffeomorphism, by adding some ad hoc boundary fields.
Then we add some observations concerning quantization, the
density of states and the entropy of this system.  

\s{Two dimensional models}

To investigate some of the properties that make black
holes so special, it is useful to study the two dimensional effective
theories that can be obtained by 
dimensional reduction from higher dimensional ones. Many interesting
scalar-tensor two dimensional theories can be obtained in this
way. One important class of theories with just one scalar and the
metric, emerges from spherically symmetric reduction. A much studied
example is the Jackiw-Teitelboim model\cite{jt84} (shortened JT
therefrom) which 
can be obtained by dimensional reduction from three-dimensional anti-de
Sitter space AdS$_3$. Theories in which one or more scalars couple to
gravity in two dimensions fall under the generic name of ``dilaton
gravity''\cite{stro-gidd:1993pr}. It must be 
recognized, however, that using these models in black hole physics one
neglects all transverse excitations of the black hole, and therefore
works with an approximate scheme that may simply be wrong (see 
\cite{Brustein:2001prl} for an alternative approach). \\   
We are interested in studying the reduction of
the Schwarzschild's black hole, so we start with a metric of the form
(we take the signature ``mostly plus'')

\beq\label{metred}
ds^2=\ga_{ab}dx^adx^b+X^2(x^a)d\om^2_{d-2}
\eeq
where $d\om_{d-2}^2$ is the line element of a compact space with
volume $\om_{d-2}$. Up to a total derivative the Einstein-Hilbert
action is

\begin{eqnarray}
I_d&=&\frac{\om_{d-2}}{16\pi G_d}\int_{\Si}
\left((d-2)(d-3)X^{d-4}\ga^{ab}\partial_aX
\partial_bX+X^{d-2}\ca R[\ga]\right.\nn \\
&&+\left.\ka(d-2)(d-3)X^{d-4}-2\La X^{d-2}\right)
\,|\ga|^{1/2}d^2x
\label{azione}
\end{eqnarray}
where $\ka=0,\pm1$ is the curvature of $d\om_{d-2}^2$ and \(\Si\) is a 
two-surface. In three dimensions we get the JT model; taking
$\om_1=2\pi$ the action is
\[
I=\frac{1}{8G}\int\,X(\ca R-2\La)|\ga|^{1/2}d^2x
\]

In four dimensions the action (\ref{azione})
is quadratic in the field\footnote{Viewed as a field in two
  dimensions, $X$ does not have canonical dimension.} $X$

\begin{eqnarray}
I_4=\frac{\om_2}{8\pi G_4}\int_{\Si}|\ga|^{1/2}
\left(\ga^{ab}\partial_aX
\partial_bX+\frac{1}{2}X^2\ca R[\ga]+\ka-\La X^2\right)
\,d^2x
\label{azione4}
\end{eqnarray}
and leads to the following field equations
\beq
-\Box_{\ga}X+X\ca R[\ga]-2\La X=0
\label{feq1}
\eeq
\beq
T_{ab}\equiv\partial_aX\partial_bX-\frac{1}{2}\ga_{ab}
\partial_cX\partial^cX-\frac{1}{2}\nabla_a\partial_b
(X^2)+\frac{1}{2}\ga_{ab}
\Box_{\ga}(X^2)+\frac{1}{2}(\La X^2-\kappa)\ga_{ab}=0\label{feq2}
\eeq 
The kinetic term of the \(X\) field in \eqref{azione4} is that of a
spacelike string coordinate, so one is tempted to think that $X$ is
not a physical field. However, when the curvature coupling is non zero
either sign for the kinetic energy is legitimate, since the $X$ field
mixes with the conformal factor of the metric in such a way that the
signature of the resulting kinetic matrix is $(+,-)$. In fact, by an
appropriate  Weyl rescaling, \(\ga_{ab}=X^{-2}q_{ab}\), one obtains
the action (we set $\ka=1$ and $\om_2=4\pi$ as is appropriate to
spherical symmetry)   

\[
I_4=-\frac{1}{2G}\int_{\Si}|q|^{1/2}\left(q^{ab}
\partial_aX\partial_bX-2^{-1}X^2\ca R[q]-X^{-2}+\La\right)\,d^2x
\]
This has now a standard kinetic term but a non trivial potential. 
Both theories are just a version of dilaton
gravity\cite{stro-gidd:1993pr, odintsov:1991, odintsov:1993,russo:92},
so are equivalent to a $c=26$ conformal sigma model with
$2D$ target space\footnote{With coordinates $(X,\phi)$, $\phi$ the
  Liouville field.}. The field equations for the action
\eqref{azione4} in a flat conformal gauge with
\(\ga_{ab}=\eta_{ab}\exp(2\phi)\) and coordinates $(t,x)\in\R^2$, are    

\begin{subequations}\label{flateq}
\begin{align}
&\Box_{\eta}X+2X\Box_{\eta}\phi+2\La Xe^{2\phi}=0\label{eq1}\\
&X\Box_{\eta}X+\eta^{ab}\partial_aX\partial_bX
=(\ka-\La X^2)e^{2\phi}\label{eq2}
\end{align}
\end{subequations}

plus two constraints (i.e. the missing equations of motion $T_{ab}=0$
written with respect to null coordinates $x_{\pm}=t\pm x$) 

\beq
T_{\pm\pm}=-X\left(\partial^2_{\pm}X+2\,\partial_{\pm}\phi\partial_{\pm}
X\right)=0 
\eeq

The remaining equation, $T_{+-}=0$, is equivalent to \eqref{eq2}. Some
solutions are:  for $\La=0$ the trivial, flat solution $\ka=1$, $X=x$,
$\phi=0$ and the non trivial Schwarzschild's solution with $\ka=1$ and
mass parameter $M=2a$ 

\[
x=X+a\ln(X/a-1), \qquad e^{2\phi}=1-a/X
\]

We see that $x$ is the well known tortoise coordinate for the
Schwarzschild black hole, while the dilaton $X$ is a non trivial
function of $x$. 
For $\La>0$ and $\ka=1$ there is a constant solution $X=1/\sqrt{\La}$,
$R[\ga]=2\La$, namely the Nariai extreme black hole $dS_2\times S^2$
with product metric. For $\La<0$ and $\ka=-1$ there is 
a Bertotti-Robinson solution $AdS_2\times S^2$ with product metric, and of
course there will be many other solutions with varying $X$.

Conversely, given a solution of two dimensional dilaton gravity we can
form the metric \eqref{metred} and this will solve Einstein equations
in higher dimensions.

The kinetic term can also be made to disappear, by using the Weyl
rescaling $\ga^{ab}\to X\ga^{ab}$ and a field reparametrization
$\eta=X^2$ at the same time. Then the action takes the form 

\begin{subequations}\label{2ddilatonaction}
\begin{align}
&I=\frac{1}{4G}\int_\Si |\ga|^{1/2}\at\eta\ca{R}[\ga]+2V(\eta)\ct
 d^2x\label{12d}\\
&V(\eta)=\frac{1}{\sqrt{\eta}}-\La\sqrt{\eta}\label{22d} 
\end{align}
\end{subequations}

and this is the action we will use. From \eqref{flateq} we see that
the dilaton mixes non trivially with the Liouville field. 
In principle, the effective action governing the $X$, or $\eta$,
fields, can be obtained by integrating $\exp iI$ over all $2D$ metrics
on $\Si$ modulo all diffeomorphisms. It is this last integration that
produces the $c=-26$ anomaly that in dilaton gravity is supposed to be
cancelled by the other fields.

So we have to study a classical two dimensional dilaton theory.
In the case of dimensional reduction from a four dimensional theory,
$V(\eta)=1/\sqrt{\eta}-\La\sqrt{\eta}$, but we shall
study more general actions with arbitrary potentials. For example,
with a linear dilaton potential one has the JT model \cite{jt84}, and
with a constant potential the CGHS model \cite{cghs92}.  
The equations of motion are:
\beq
\ca{R}[\ga]+2\pa_\eta V(\eta)=0,\qquad
\ga_{ab}\Box\eta-\nabla_a\nabla_b\eta=V(\eta)\ga_{ab}
\eeq
Note that the value of the dilaton on the horizon of the parent,
higher dimensional black hole, is the area of the horizon
spatial section divided by $4\pi$. In this way the $2D$ theory
remembers its higher dimensional origin.

This is a good place to mention an important consistency 
condition, which is that the entropy of the original higher
dimensional black hole must match correctly with the entropy of the
two dimensional theory. In higher dimensions and within general
relativity (i.e. with the Einstein-Hilbert action determining
dynamics) the entropy is given by the usual Bekenstein-Hawking area
law, $S=A/4G$. In two dimensions we do not have horizons with finite
area but we have a valid substitute, the Noether
charge\cite{wald:93}. Given a stationary black hole solution 
in the theory \eqref{2ddilatonaction}, with Killing vector field
$\xi^a$ and surface gravity $\ka=-2^{-1}\ep^{ab}\nabla_a\xi_b$,
one can define a conserved charge on the horizon that is  
quite independent on the form of the potential. This Noether charge
$0$-form\footnote{The Noether charge is a $(d-2)$-form in $d$
  dimensions.} turns out to be 
\beq\label{Q0}
Q_H=\ka\eta_0/2G=\ka A/8\pi G
\eeq
where $\eta_0=A/4\pi$ is the dilaton restricted on the horizon. Given
the charge, the entropy is\cite{wald:94}   
\[
S=2\pi\ka^{-1}Q_H=\pi G^{-1}\eta_0=A/4G
\] 
as we wanted to show. Even if there were a mathematical explanation of
this coincidence it remains a little bit surprising. In the event that
a state counting in $2D$ dilaton gravity succeeds in computing the
Noether charge, that would be a successful state counting for the
entropy of almost all higher dimensional, non rotating black holes!\\   
In the following, we shall investigate the Hamiltonian structure of these
generic dilatonic theories\footnote{A rather general treatment of
phase space formulation of $2D$ gravity models can be found in
\cite{krv:1997pr}. The quantum theory is studied in
\cite{stro-gidd:1993pr}. See also \cite{strobl:00,odintsov:2000,grumiller:2002} for extensive reviews on the
subject.}.   
  
\s{The Hamiltonian generators and the central charge}

In the first part of this section we briefly summarize the classical
Hamiltonian for two dimensional dilaton gravity. Then we shall 
directly compute the 
Poisson bracket of the Hamiltonian generators in the presence of
boundary terms. This is inspired by the work of Brown, Lau  and York
\cite{brownlauyork00}, who computed the Poisson brackets of the
Hamiltonian generators with boundary terms, in four dimensions.  
The ADM form of a two dimensional metric is 

\beq\label{2dmetric}
ds^2=-N^2dt^2+\sigma^2(dx+N^xdt)^2
\eeq

and we take the action in the form given by \eqref{2ddilatonaction}
, with a general potential. Discarding for the moment the boundary
terms, the Lagrangian density $\ca{L}$ becomes (we set $2G=1$ for the
time being and restore Newton constant afterwards) 
\beq
\ca{L}= -\frac{\dot{\eta}\dot{\sigma}}{N}-N\left(\frac{\eta
'}{\sigma}\right)'+{\sigma
N}V(\eta)-\frac{N^x(N^x\si)'\eta'}{N}+\frac{\dot\sigma N^x\eta'}{N}
+\frac{\dot\eta(\si N^x)'}{N}
\eeq
where we agree that one or more primes will denote spatial derivatives
(i.e. derivatives with respect to $x$) and one or more overdots will
indicate time derivatives. 
If we want a Hamiltonian formulation we also need to find the
conjugate momenta $\Pi_\eta$, $\Pi_\sigma$ and $\Pi_N$, which are
\beq\label{moment}
\Pi_\eta=\frac{-\dot{\sigma}+(N^x\sigma)'}{N},\qquad
\Pi_\sigma=\frac{-\dot{\eta}+N^x\eta'}{N}, \qquad \Pi_N=0
\eeq
The bulk Hamiltonian will be a sum of constraints
\beq
H=\int dx(N\ca{H}_{\perp}+N^x\ca{H}_x)
\eeq
respectively given by
\beq\label{const}
\ca{H}_{\perp}=-\Pi_\eta\Pi_\sigma+\at\frac{\eta'}{\sigma}\ct'-V(\eta)\sigma
\sim 0,\qquad
\ca{H}_x=\Pi_\eta\eta'-\sigma\Pi_\sigma'\sim 0
\eeq
The constraint functionals can also be smeared with test fields
$\xi^{\mu}$ and used as generators of the gauge
symmetry, which in our case is the diffeomorphism symmetry generated
by the vector field $\xi^{\mu}$.\\
To obtain well defined generators the test fields $\xi^{\mu}$, and
possibly some first order derivatives as well, must vanish on the
boundary. So we define the smeared Hamiltonian constraints
\beq
H[N^x]=\int_\Si dx N^x\ca{H}_x,\qquad H[N]=\int_\Si dx N\ca{H}_{\perp}
\eeq
and $H=H[N,N^x]=H[N]+H[N^x]$. There are terms in the Hamiltonian giving
origin to boundary terms in the course of variations. Looking at
\eqref{const}, these terms are $N\at\frac{\eta'}{\sigma}\ct'$ and 
$N^x\at \Pi_\eta\eta'- \si\Pi_\si'\ct$. So, collecting the variations 

\beq\label{hvar}
\de H=\aq N\frac{\de\eta'}{\si}
-N\frac{\eta'}{\si^2}\de\si
-N'\frac{\de\eta}{\si}+N^x\Pi_\eta\de\eta-N^x\si\de\Pi_\si\cq_{\pa\Si}+
\dots
\eeq

the dots denoting bulk terms giving the equations of motion.
This expression shows that the action functional is not differentiable
if $N\neq 0$ or $\si^{-1}N'\neq 0$ or $N^x\neq 0$ on the boundary.
To restore differentiability we have to add suitable boundary terms to 
the Hamiltonian generators\cite{ReTe}, but these terms will depend on
the choice of 
boundary conditions, i.e. on the physics we are going to describe, and the
physics we have in mind is the one describing the presence of a static
black hole from the viewpoint of an observer who chooses to stay
permanently outside it. In that case the boundary term \eqref{hvar}
has an inner contribution at the horizon, and therefore we will need
to impose (rather specific) boundary conditions on the metric near the
horizon.

First, the $U(1)$ isometry
corresponding to time translations of the metric has a fixed point set
at the horizon position $x_0$, so $N(x_0)=0$. Second, experience with 
black holes requires that
the geometry of the metric \eqref{2dmetric} be isometric to a flat
disk\footnote{More correctly, this is true 
after a Wick rotation $\tau=-it$. In a Lorentzian world, the isometry
is with a flat two-dimensional Rindler space. For extreme states, the
isometry is with two-dimensional anti-de Sitter space.}. Denoting
$\ka$ the surface gravity of the black hole, the Euler 
characteristic of the metric \eqref{2dmetric} is found to be
$\chi=(\ka\si)^{-1}N^{'}$, so requiring $\chi=1$ gives the regularity
condition  

\beq
(\si^{-1}N^{'})_{|x=x_0}=\kappa 
\label{euler}
\eeq

For extreme states in particular (solutions with $\ka=0$),
(\ref{euler}) is replaced with $N^{'}(x_0)=0$, because in this
case the black hole has the topology of the annulus.\\
Finally, it was noted in \cite{bcmmwy90} that 
the Euler characteristic of the full black hole metric must be
two\footnote{Because \(\chi(black hole)=\chi(disk)\times\chi(sphere)\)
  and \(\chi(sphere)=2\).}, which gives the third condition
$\si^{-1}(x_0)=0$.

For the metric \eqref{2dmetric} we summarize these conditions in the
form 

\begin{subequations}\label{bcblack}
\begin{align}
&N(x_0)=0\label{bc1}\\
&\si(x_0)^{-1}=0\label{bc2}\\
&(\si^{-1}N^{'})_{|x=x_0}=\kappa,\;\,\,\,\,\textrm{a fixed non
  zero constant}\label{bc3}
\end{align}
\end{subequations}

The modified Hamiltonian generators which are differentiables under
the given boundary conditions can be obtained by looking to
Eq.~\eqref{hvar}, and are\footnote{We omit the terms at
  infinity which are familiar and well known.}

\begin{subequations}\label{hs}
\begin{align}
H[N^x]&=\int_\Si\,dx N^x\ca{H}_x-(N^x\at
\Pi_\eta\eta-\Pi_\sigma\sigma\ct)|_{\pa \Si}\label{hs1}\\
H[N]&=\int_\Si dx N\ca{H}-\left. N
\frac{\eta'}{\sigma}\right|_{\pa\Si}
+\left. N'\frac{\eta}{\sigma}\right|_{\pa \Si}\label{hs2}
\end{align}
\end{subequations}
where $\partial\Si$ is the surface $x=x_0$ (a point). 
One can verify that the variations on the surface $x=x_0$
are all zero. The modified Hamiltonian generators are in fact
observables because they commute with the equations of motion as well 
as the constraints\cite{balachandran}.

Now that we have a well defined Hamiltonian system, we may evaluate
$\dot\Pi_\sigma$ and $\dot\Pi_\eta$
\beq
\dot\Pi_\sigma=\Pi_\sigma'N^x-\frac{\eta'N'}{\sigma^2}+V(\eta)N, \quad
\dot\Pi_\eta=\at\Pi_\eta N^x\ct'-\at\frac{N'}{\sigma}\ct'+\sigma
N\frac{\pa V(\eta)}{\pa\eta}
\eeq 
Together with \eqref{moment}, these are equivalent to the original
Lagrangian field equations.

The symmetries of the system associated with the vector $\xi^\mu$ are
generated by the functionals $\ca{O}(\xi)$, where 
\beq
\ca{O}[\xi]=\int dx\at\xi^\o\ca{H}+\xi^\p\ca{H}_x\ct +Q[\xi]
\eeq
and $\xi^\o=N\xi^t$, $\xi^\p=\xi^x+N^x\xi^t$ and $Q[\xi]$ are the
boundary term introduced above. These generators are attached to the
boundary in the sense that vector fields $\xi$, $\chi$, which
on the boundary coincide together with their first derivatives, define
generators differing by a constraint.

Now computing the Poisson bracket
between two generators we obtain (with a substantial effort) a result
of the form  
\beq\label{algop}
\ag\ca{O}{[\xi]},\ca{O}[\psi]\cg=\ca{O}[[\xi,\psi]_{SD}]+K[\xi,\psi]
\eeq
where $K[\xi,\psi]$ is an a priori non vanishing central charge
\beq\label{ccharge}
K[\xi,\psi]&=&\at [\xi,\psi]_{SD}^\p\Pi_\eta\eta
-(\xi^\p{\psi^\p}'-{\xi^\p}'\psi^\p)\Pi_\sigma\sigma\ct_{|\pa \Si}\\
&+&\at(\xi^\p\psi^\o-\psi^\p\xi^\o)\at\Pi_\eta\Pi_\sigma+V(\eta)\sigma\ct-
\frac{\eta}{\sigma}([\xi,\psi]_{SD}^\o)'\ct_{|\pa \Si}\nn
\eeq
and the bracket $[\,,\,]_{SD}$ is 
the well known surface deformation algebra for a two dimensional theory  
\beq\label{sdef}
[\xi,\psi]_{SD}^\o=\xi^\p{\psi^\o}'-{\psi^\p\xi^\o}',\qquad
[\xi,\psi]_{SD}^\p=\frac{\xi^\o{\psi^\o}'-\psi^\o{\xi^\o}'}{\si^2}+
\xi^\p{\psi^\p}'-\psi^\p{\xi^\p}'
\eeq
We see that the possibility of a central charge is real. Indeed, as is
well known\cite{BrownHenn}, in
three dimensional anti-de Sitter gravity the boundary at infinity
gives a non vanishing $K[\xi,\psi]$. As we noted above, if there is a
black hole and we use the Hamiltonian of a static external observer,
then there is a contribution to \eqref{ccharge} coming from the inner
boundary at the horizon.
But in the next section we shall show that if we impose the presence
of a black hole, this contribution to the central charge actually
vanishes.  

\s{The black hole and the near horizon symmetry}

As far as we know, the presence of a black hole is always associated
to the existence of an event horizon. Such an horizon is not a real,
physical 
boundary of space-time, but from the viewpoint of an observer who
chooses to stay permanently outside the black hole, we may treat it
like a boundary. The horizon boundary conditions \eqref{bcblack} then makes
it possible to predict the exterior region alone, by giving initial data
on a partial Cauchy surface extending from the horizon to spatial
infinity, always external to the black hole. This is just the
property of asymptotics predictability, a cornerstone of classical
black hole physics.    

If we analyze the behaviour of the  Hamiltonian generators of our
theory, we find that there are some transformations which change these 
boundary conditions and there are other transformations leaving them
unchanged: these are the ``gauge'' transformations of the theory. The
other ones change the black hole. 
We may fix $N^x=0$ since on static solutions it can be globally
removed by a change of coordinates. We find that
${\Pi_\eta}_{|\pa\Si}={\Pi_\si}_{|\pa\Si}=0$ and the horizon central
charge \eqref{ccharge} on such static solutions becomes
\beq\label{boundtermonshell}
K[\xi,\psi]=
\at(\xi^\p\psi^\o-\psi^\p\xi^\o)\at\frac{\eta'}{\si}\ct'-
\frac{\eta}{\sigma}([\xi,\psi]_{SD}^\o)'\ct_{\pa \Si}
\eeq
We are interesting in studying the class of diffeomorphisms $\xi$ that do not
change the black hole, in particular we want that the boundary conditions
\eqref{bcblack} be preserved. So near the horizon at $x_0$ we ask for the
following asymptotic behaviour of the lapse function
\beq
N^2=O(x-x_0), \quad N'/\si=\ka+O(x-x_0)\;\;\;\textrm{as}\;\;\;x\to x_0 
\eeq

where the symbols $\phi=O(\psi)$ means that the ratio $\phi\!:\!\psi$
is bounded as $x\to x_0$.
The linear dependence on $x-x_0$ is necessary to have a non zero
surface gravity. More precisely, the regularity conditions
\eqref{bcblack} require the near horizon expansion 

\begin{subequations}\label{grup}
\begin{align}
N^2&=(-g^{tt})^{-1}=4\ka^2x_0(x-x_0)+\al(t)(x-x_0)^2+O_3\label{prima}\\
\si^2&=g_{xx}=\frac{x_0}{x-x_0}+\Phi(x,t)\label{seconda}\\
\ga_{tx}&=\la(x-x_0)+O_3\label{terza}
\end{align}
\end{subequations}
where \(\Phi\) and its first derivatives are regular at the
horizon. The constant $x_0$ is arbitrary at this stage, but can be
fixed in terms of the mass at infinity. For the Schwarzschild solution
one finds of course $x_0=1/2\ka$. A vector fields $\xi$
preserving these conditions can at most change the
metric according to the rules
\[
\de N^2=\ca{L}_\xi N^2=O(x-x_0), \qquad \ca{L}_\xi\si^2=O(1),
\qquad \ca{L}_\xi\ga_{tx}=O(x-x_0)
\]

and then we will have well defined observables
$\ca{O}[\xi]$ with boundary charges $Q[\xi]$ (we restore $2G$ now)    
\beq\label{bobs}
Q[\xi]=\left. \frac{\eta\,{\xi^\o}'}{2G\sigma}\right|_{\pa \Si}
\eeq
These boundary observables (in the sense explained in
Ref.~\cite{balachandran}) 
are the Noether charges belonging to the diffeomorphism generated by
$\xi$. If $\xi$ is the Killing time translation symmetry then
$\xi^\o=N$ and the boundary conditions \eqref{bcblack} gives
$Q=Q_H=\ka\eta_0/2G$ (cf. \eqref{Q0}).\\ 
We recall that the boundary conditions are essential to have well
defined, i.e. differentiable, Hamiltonian generators
$\ca{O}[\xi]$. The diffeomorphism which preserve the conditions 
\eqref{bcblack} lead to well defined Hamiltonian generators. If we
drop the condition $\de\si^2=O(1)$ or $\de N^2=O(x-x_0)$, the 
Hamiltonian generators are not differentiable, and consequently, 
not well defined as generators. 

The regularity conditions being equivalent to the
expansions \eqref{prima}, (\ref{seconda}), \eqref{terza} near the
bolt, an elementary computation gives the components of the
diffeomorphism preserving generators as  
\beq
\xi^x=-\ka^{-1}N^2\partial_t\xi^t+O(N^3), \qquad \xi^t=O(1)
\label{dif}
\eeq
\beq
\xi^x\frac{1}{(x-x_0)^2}-2\partial_x\xi^x\,\frac{1}{x-x_0}=O(1)
\label{dif1}
\eeq

Vector fields satisfying these equations and having limit on the
horizon together with their first derivatives, have two important
properties. First, given
$\xi_1$, \(\xi_2\) both satisfying (\ref{dif}), the Lie bracket also
satisfies (\ref{dif}), so these vector fields close a Lie algebra.\\
The second property requires the surface deformation
algebra \cite{Teit}, which has the form 
\beq
[\xi_1,\xi_2]_{SD}^{\perp}=\xi_1^x\partial_x(N\xi_2^{\tau})-\xi_2^x
\partial_x(N\xi_1^{\tau})
\eeq
\beq
[\xi_1,\xi_2]_{SD}^x=\xi_1^x\partial_x\xi_2^x-\xi_2^x\partial_x\xi_1^x+
N^3\left(\xi_1^{\tau}\partial_x(
N\xi_2^{\tau})-\xi_2^{\tau}\partial_x(N\xi_1^{\tau})\right) 
\eeq 
Then, if $\xi_1$ and \(\xi_2\) satisfy (\ref{dif}),
\[
[\xi_1,\xi_2]_{SD}=[\xi_1,\xi_2]+O(N^3)
\]
Requiring the  vector fields $\xi$ together with their first
 derivatives to have a limit on the horizon, the equations
 \eqref{dif}, \eqref{dif1} imply diffeomorphisms of the form 
\begin{subequations}\label{1e2}
\begin{align}
\xi^t&=\chi_0+\chi_1(x-x_0)+\sum_{k=2}^{\ii}a_k(t)(x-x_0)^k\label{uno}\\
\xi^x&=\sum_{k=1}^{\ii}b_k(t)(x-x_0)^{k+1}\label{due}
\end{align}
\end{subequations}
where the functions $b_k(t)$ are proportional to the time derivatives
of the functions $a_k(t)$. These vectors form a Lie subalgebra of the
Lie algebra of all vector fields. Using the asymptotic behaviour of
$\xi^a$ on the horizon it is a simple matter to show that
$K[\xi_1,\xi_2]$, as given by \eqref{ccharge}, vanishes. Moreover the
boundary observables \eqref{bobs} are all proportional to the Noether
charge $Q_H=\ka\eta_0/2G$, and form therefore a one-dimensional abelian
algebra.

We recall that the Poisson bracket of two differentiable generators is a
differentiable generator \cite{brownhenn86}, so the observable
$\ca{O}[\xi,\psi]_{SD}$ automatically includes the correct boundary
terms up to a constant $K[\xi,\psi]$ which depends only on the
asymptotic form of the vectors $\xi$, $\psi$. What happens here is
that this constant actually vanishes for every choice
of $\xi$ and $\psi$ satisfying our boundary conditions, and that the
bracket $[\xi,\psi]_{SD}$ is a pure gauge, i.e. it has zero charge.

We also recall that the Hamiltonian generators are defined up to
constants. Such constants are usually fixed by normalizing the
Hamiltonians to given values in some background manifold. Our
charges are normalized so that $Q_H$, given by \eqref{Q0}, is the
mass as measured at infinity.  
Sometimes other normalizations are proposed. In \cite{silva02}, for
example, normalized to zero values are the horizon 
charges of the generators of the enveloping algebra
$\ca{O}[\hat\xi_{-1}]$, $\ca{O}[\hat\xi_0]$, $\ca{O}[\hat\xi_1]$, and
after this a central charge $K[\hat\xi_{-2},\hat\xi_2]$ is found. 
We stress that the normalization cannot determine whether a central
charge is present or not, it can only affect its actual value.

The only admissible way to find a non zero central charge on the
horizon is to 
stretch the horizon and consider diffeomorphisms that do not change the
boundary value of the Hamiltonian on this ``stretched'' horizon
\cite{carlip02,silva02}. This calculation leads to nonzero charges and
to a nonzero central charge when the stretched horizon tends to the physical
horizon. In this cases the diffeomorphisms are not defined in the limit.
This seems to be unsatisfactory for the definition of a conformal algebra on
the horizon. Finally, there remains the possibility that the symmetries
involved in the boundary observables are the transverse angular
vectors that we lost because of our dimensional reduction. This seems
unlikely to us, since these diffeomorphisms should preserve the area of
the horizon and there are no central extensions for the area preserving
diffeomorphisms on the sphere\cite{bars:88}.

\s{Another route to a central charge}

The fact that no central charge can emerge from boundary observables
does not prove the absolute absence of a classical central charge in
$2D$ theories of the kind we have discussed. There remains the
possibility of a charge emerging from the bulk observables, as a
Schwinger term in the constraint algebra. In fact,
it is well known that in two dimensions special conditions affect
the constraint algebra. One fact is that the Hamiltonian generators may  
not vanish on shell, but the equations of motion can still be generally
covariant. This is possible because in two dimensions the structure
coefficients of the constraint algebra (not really an algebra)    
\[
\{H_{\mu}(x),H_{\nu}(x^{'})\}=\int\,dx^{''}\,K_{\mu\nu}^{\rho}(x,x^{'};x^{''})
H_{\rho}(x^{''})
\]
can be made independent on the canonical variables. With
generators  $H_{\mu}(x)$ not constrained to vanish, a central term
can be consistently added to the left hand side of this equation. 

In our case however, we started with a covariant theory and this has
the consequence that the generators must all vanish on shell, as is well
known. No classical Schwinger terms are then possible in the
constraint algebra. We think this is one reason why people searched
for a central extension in the algebra of boundary observables, rather
than among constraints.

We now present an argument, similar in spirit to Verlinde's rewriting
of FRW equations as an entropy formula\cite{verlinde}, which suggest
the presence of 
a classical central charge. The Hamiltonians \eqref{hs} were deduced
for a singular foliation where all surfaces intersect at the
bifurcation point of the event horizon. They consist of a term at
infinity, giving the energy, and a term at the horizon. As we
explained, this term is what fixes the disk topology of the black
hole. Suppose we write this Hamiltonian in the form given by a
conformal field theory on the disk\footnote{The factor
  $\ka^{-1}=\be/2\pi$ means that scaling the period of euclidean time
  from $2\pi$ to $\be$ changes the Hamiltonian as written.} 

\beq\label{hamic}
\ka^{-1}H=L_0+\bar{L}_0-\frac{c+\bar{c}}{24}
\eeq

where the first two terms correspond to the boundary at infinity (the
energy) and the last two to the boundary at the horizon. For a non
singular foliation, with no intersections at the horizon, there would
be no such terms in the Hamiltonian. This would correspond to the
Hamiltonian of a conformal field theory as given on the cylinder. Let we
assume a symmetric contribution of left/right moving modes,
i.e. $L_0=\bar{L}_0$ and $c=\bar{c}$ (non rotating black holes). The
mass at infinity is $\ka A/4\pi G$ so from \eqref{hamic} we
get\footnote{Recall that the dilaton at the horizon is
  $\eta_0=A/4\pi$.}  
\beq\label{vir}
L_0=\frac{A}{8\pi G}=\frac{\eta_0}{2G}
\eeq

The Noether charge at the horizon is $\ka\eta_0/2G$, so again from
\eqref{hamic} we get
\beq\label{c}
c=\frac{6\eta_0}{G}=\frac{3A}{2\pi G}
\eeq

Since $L_0$ is comparable to $c$, the entropy should be\cite{cardy}
\beq\label{cardy}
S=2\pi\sqrt{\frac{c}{6}\at
  L_0-\frac{c}{24}\ct}=\frac{\pi\eta_0}{G}=\frac{A}{4G} 
\eeq

Eq.~\eqref{c} is the memory of dimensional reduction. It can be
represented pictorially as follows: since the actual geometry is the
disk times a sphere with area $A$, we have in effect a $2D$ theory
attached at each Planck-sized cell on the sphere. The central charge
is additive, whence the result.

There is also another interpretation of \eqref{hamic}, which is that
it can be 
viewed as an Euclidean conformal field theory on the horizon. 
In fact, as shown in \cite{mopi}, it is possible to define a free
quantum field theory on the future and past horizons in terms of $2D$
conformal field theories. 
These $2D$ fields are: in the holomorphic case $\hat{\phi}(z)=\sum_n
z^na_n/n$ and in the anti-holomorphic case $\hat{\phi}(\bar{z})=\sum_n
\bar{z}^nb_n/n$. 
Here $z=\exp(-i \arctan(\ka\, it+\ka\,r_*) )$, where $r_*$ is the
tortoise coordinate, and $t$ is the complexified time variable on
the horizon as seen by an observer at spatial infinity\footnote{See
  \cite{mopi} for details.}.  
As usual, the Virasoro generators are the Fourier component of the
Euclidean stress tensor  
$T(z)=\sum_{\,n\in\mathbb{Z}/\{0\}}{z^{-n-2}{L_n}}$,
$T(\bar{z})=\sum_{\,n\in\mathbb{Z}/\{0\}} {\bar{z}^{-n-2}{\bar{L}_n}}$.
Since the central charge of this theory is different from $1$, it
can be naturally interpreted as a two dimensional linear dilaton conformal
field theory whose stress-energy tensor has the following components
\cite{polchinski}
\beq
T_{zz}=T(z)&=& -:\pa_z\hat{\phi}(z)\,\pa_z\hat{\phi}(z):+\,V
\,\pa_z^{\,2}\,\hat{\phi}(z),\\
T_{\bar{z}\bar{z}}=T(\overline{z})&=&
-:\pa_{\bar{z}}\hat{\phi}\,(\bar{z})\,\pa_{\bar{z}}\hat{\phi}(\bar{z}):+\,
V\,\pa_{\bar{z}}^{\,2}\,\hat{\phi}(\bar{z}).
\eeq
with $V^2=\eta_0/G=A/(4\pi G)$. The corresponding central charge $c$
is equal to $1+6\eta_0/G$. In the semi-classical regime 
$\eta_0/G>>1$, so this does not affect the computation of the entropy in
\eqref{cardy}. Moreover, as shown in \cite{mopi}, the field
$\hat{\phi}(z)$ can be interpreted as a free field in the bulk of a
Rindler spacetime.

We think this qualitative argument is suggestive enough to deserve
a more detailed investigation. We note that with the usual
normalization of the action that is used in the JT model of $2D$ gravity,
corresponding here to $G=1/2$, the central charge $c=12\eta_0$ is what
has been found in several researches on the JT model with anti-de
Sitter boundary conditions\cite{CadMin,klem,VanzCaldCate,catvanz}. The
present argument suggests that the central charge is universal and
independent on the form of the dilaton potential $V(\eta)$. 

In the next section we abandon the search for a central charge, and
present a very tentative model suitable to describe the quantum hairs.
Our exercise follows a suggestion made in \cite{balachandran}.


\s{A covariant Action}

The theory of relativity in a domain without boundary is invariant
under diffeomorphisms, so diffeomorphic metrics will describe the
same physics. In this sense the diffeomorphisms are like gauge
transformations. 
The situation is totally different in the presence of a boundary, for in
this case the gauge group is smaller. It contains only
transformations which do not change the boundary value of the action. 
Technically speaking, this means
that the generators of these transformations are the Hamiltonians
corresponding to prescribed boundary conditions.

If we insist to consider the horizon as a boundary we necessarily
lose part of the  gauge group, and there will be gauge
transformations which are not symmetries of the black hole.
But we expect physics to be invariant under the full gauge
group. There is a way to restore the gauge invariance of such
systems\cite{balachandran}, and this is adding suitable boundary field
to the action, possibly coupled with the bulk fields.

We have analyzed the generators of the permitted transformations in
Sec.~[3]. Some of them are pure gauge transformation ($N=0$, $N'=0$
and $N^x=0$ on the boundary). In this case the Hamiltonian generators
$H[N,N^x]$ are like ``Gauss laws'', and the commutator of such
Hamiltonians vanishes on shell.  
There are other generators that commute with the constraints,
but they are not zero on shell, they are observables.
To restore the gauge invariance we will add new external fields
living on the horizon. We take the gravitational part of the action in
the form 
\beq
I=C\int_\Si |g|^{1/2}\at \frac{1}{2}\eta\ca{R}[g]+V(\eta)\ct d^2x
\eeq
with $V(\eta)=\frac{1}{\sqrt{\eta}}$ in the case of an asymptotically
flat four dimensional theory. 
The variation of this action produces the boundary terms we discussed
above. Let us consider for simplicity only the boundary term of a
diagonal two dimensional metric, $(-N^2,\sigma^2)$, with fixed
horizon at $x=x_0$, so $N(x_0)=0$. Under the small change
\beq
N\to N+\delta N,\qquad\quad
\sigma\to \sigma+\delta\sigma,\qquad\quad
\eta\to \eta+\de\eta
\eeq
new boundary terms will appear in the action variation, making it not
invariant 
\beq
\de I=\dots +\int
 dt\aq-N\de\at\frac{\eta'}{\si}\ct+\at\frac{N'}{\si}\ct\de
\eta-(N^x\Pi_\eta)\de\eta+(N^x\si)\de\Pi_\si\cq_{\pa\Si}    
\eeq

Only diffeomorphisms not bringing boundary terms are allowed
as invariances of the action, but the gauge group should be
larger. To obtain a gauge invariant action while keeping the
boundary, we add the following extra fields $\phi_1$, $\phi_2$,
$\phi_3$ on the boundary, and assume the transformation rules
\beq
\phi_1\to \phi_1- \de\frac{\eta'}{\si}_{|0} \qquad \phi_2\to
\phi_2-\de\eta_{|0} \qquad \phi_3\to \phi_3-\de(\Pi_\si)_{|0}
\eeq
Then adding the term 
\beq
I_A=\int\,dt\at N\phi_1-\at\frac{N'}{\si}-N^x\Pi_\eta \ct\phi_2-
N^x\si\phi_3\ct_{x\to x_0}
\eeq
the action $I+I_A$ becomes gauge invariant under the full
diffeomorphism group. 
At this level the new fields have no
dynamics. But nothing forbid us to add a gauge invariant dynamical
part in the action with this covariant derivative
\beq
D\phi_1= \pa\phi_1+\pa \at\frac{\eta'}{\si}\ct_{|0} \qquad D\phi_2=
\pa\phi_2+\pa \eta_{|0} \qquad D\phi_3=\pa\phi_3+\pa{\Pi_\si}_{|0}
\eeq
so we may add the kinetic term
\beq
I_D=\int\La_{AB}D\phi ^AD\phi^B
\eeq
for some field independent matrix $\La_{AB}$. The new action $I+I_A+I_D$
is gauge invariant under all diffeomorphisms.
It is interesting to note that the boundary part of the action does
not depend on the form of the dilaton potential. 

Now the exercise would be to do the statistical mechanics of the
boundary fields. The strategy we wish to follow is considering the
gravitational part of the action like that of a reservoir, and the
part with the fields $\phi_A$ like the actual system.
We observe that if we are interested in transformations made at fixed
temperature $\ka$, the interaction is particularly
simple. In fact it becomes $I_A=\int(N\phi_1-\ka\phi_2)$, but the
first part is small enough that we may drop it.

Then the action of this simplified model reads
\beq\label{azionesemplice}
I_D+I_A=\int dt\at\La\, D\phi D\phi-\ka\phi  \ct_{x\to 0}
\eeq
where now $\phi$ stands for the field $\phi_2$ described above.
For the statistical mechanics of this model we need 
the Hamiltonian, $H_B=\Pi_\phi^2/\La+\ka\phi$, derived
from (\ref{azionesemplice}). One may notice that it describes a
freely falling particle with configuration variable $\phi$ in a theory
with linear potential. 
Moreover, since $\phi$ can take only positive values, we have to
introduce an  infinite barrier at $\phi=0$ in the potential of our model.
As just pointed out $\phi$ is not a
field but a quantum variable of our system, therefore we are dealing
with a problem in ordinary quantum mechanics. 

The Hamiltonian $H_B$ has a discrete non degenerate spectrum.
Though we lack an analytic expression for the energy eigenvalues
of $H_B$, it is possible to estimate their values in the high energy
limit.  
The wave function that solves the eigenvalue equation, $H\, \psi=E\,
\psi$, reads 
\beq
\psi(x)=A\Phi\at \ka ^{1/3}\, \La^{2/3}\at x-\frac{E}{\ka}\ct\ct,
\eeq
where $\Phi (x)$ is an appropriate Airy function. 
In particular, in the high energy limit the energy levels are
proportional to $n^{2/3}$. To compute the partition function we also
need to consider the density of states

\beq\label{den}
\rho(E)=\frac{\La}{\ka\pi}\sqrt{E}.
\eeq

Assuming that the black hole is described by several non interacting
bosons forming a canonical system, we can do the statistical mechanics
of this model. 

First we fix the constant $\La$ in (\ref{den}) so that the
expected energy in a canonical ensemble with temperature
$\ka/2\pi$ and density of states \eqref{den}, be equal to the mass
of the black hole, $M=\ka\,\eta_0$. This gives

\beq
\La=\frac{16}{3}\frac{\pi^{5/2}}{\zeta{(5/2)}}\ \eta_0\ \sqrt{\beta} 
\eeq

where $\zeta(5/2)\simeq1.3414$ is the Riemann zeta function at $5/2$, 
and $\beta=2\pi\ka^{-1}$ is the inverse of the Hawking temperature.
With this curious choice of $\La$ the canonical partition
function is  
\beq\label{zeta}
\log\ca{Z}\,\simeq-\int_0^{\infty} \log{\at 1-e^{-\beta E}\ct\rho (E)\, dE}=
\frac{4}{3}\pi \eta_0,
\eeq

and 
\beq
\langle E \rangle
\,\simeq\int_0^{\infty}\frac{E\,\rho(E)\,dE}{e^{\,\beta\,
    E}-1}=\frac{2\pi \eta_0}{\beta}=M 
\eeq
Moreover, the statistical entropy becomes
\beq
S=-\beta^2\frac{\pa}{\pa \beta}\frac{\log \ca{Z}}{\beta}=\frac{4\pi}{3} \eta_0
\eeq

which is a little higher ($4/3$) than the Bekenstein-Hawking value,
but satisfies the area law. Perhaps the right factor can be restored
by considering the other fields $\phi_1$ and $\phi_3$, but our point
was not so much to be exact with the entropy. Rather, we wanted to
show that the boundary fields, as suggested by the Hamiltonian
description of a diffeomorphism invariant theory, could really do the
right job.

\s{Conclusion}

It is interesting to note that the analysis of the Hamiltonian
generators near the horizon does not reproduce the microstates of black
holes. On the other hand, the relevant observables (perhaps not all)
seem localized precisely at the horizon, but only for those observers
staying permanently outside the black hole and detecting a static
field. Others, Kruskal-like observers do not need boundary observables
to give a Hamiltonian description of space-time. 
 If one tries to effectively describe the black hole, he should realize 
at a quantum level a connection between the inner and the
outer part of the horizon, through the addition of boundary
fields. These are necessary to restore the full invariance group,
broken by the choice of boundary conditions. 
We have shown that the statistical description of these fields is at
least compatible with the thermodynamics description of the black
hole.  

\section*{Acknowledgments}
We thanks Valter Moretti for useful discussions and technical 
suggestions.

\end{document}